\documentclass[journal=jacsat,manuscript=article]{achemso}
\setkeys{acs}{keywords=true}
\usepackage{float}
\usepackage{xcolor}
\usepackage[normalem]{ulem}
\usepackage{chemformula} 
\usepackage[T1]{fontenc} 

\author{Zhiyi Hu}
\affiliation{Institute of Quantum Sensing and College of Optical Science and Engineering,  Zhejiang University, Hangzhou, 310027. China}
\alsoaffiliation{School of Microelectronics, Hefei University of Technology, Hefei, 230009, China}

\author{Fengjian Jiang}
\affiliation{School of Information Engineering, Huangshan University, Huangshan, 245041, China}

\author{Jingyan He}
\affiliation{Institute of Quantum Sensing and College of Optical Science and Engineering,  Zhejiang University, Hangzhou, 310027. China}

\author{Yulin Dai}
\affiliation{Institute of Quantum Sensing and College of Optical Science and Engineering,  Zhejiang University, Hangzhou, 310027. China}

\author{Ya Wang}
\email{ywustc@ustc.edu.cn}
\affiliation{CAS Key Laboratory of Microscale Magnetic Resonance and School of Physical Sciences, University of Science and Technology of China, Hefei 230026, China}

\author{Nanyang Xu}
\email{nyxu.physics@zju.edu.cn}
\affiliation{Institute of Quantum Sensing and College of Optical Science and Engineering,  Zhejiang University, Hangzhou, 310027. China}

\author{Jiangfeng Du}
\affiliation{Institute of Quantum Sensing and College of Optical Science and Engineering,  Zhejiang University, Hangzhou, 310027. China}

\title{Four-order power reduction in nanoscale electron-nuclear double resonance with a nitrogen-vacancy center in diamond}

\abbreviations{IR,NMR,UV}
\keywords{NV center in diamond, nuclear magnetic resonance, weakly-coupled nuclear spins, phase modulation, Hartmann-Hahn condition}

\begin{tocentry}
\includegraphics{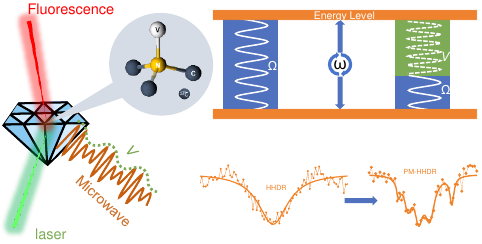}
\end{tocentry}

\begin{document}

\begin{abstract}
Detecting nuclear spins using single Nitrogen-Vacancy (NV) centers is of particular importance in nano-scale science and engineering, but often suffers from the heating effect of microwave fields for spin manipulation, especially under high magnetic fields. Here, we realize an energy-efficient nano-scale nuclear-spin detection using a phase-modulation electron-nuclear double resonance scheme. The microwave field can be reduced to 1/250 of previous requirements and the corresponding power is over four orders lower. Meanwhile, the microwave-induced broadening to the line-width of the spectroscopy is significantly canceled and we achieve a nuclear-spin spectrum with a resolution down to 2.1 kHz under a magnetic field at 1840 Gs. The spectral resolution can be further improved by upgrading the experimental control precision. This scheme can also be used in sensing microwave fields and extended to a wide range of applications in the future.
\end{abstract}

Nuclear magnetic resonance (NMR) spectroscopy is one of the most important analytical technique that is widely used in physics, chemistry, biology and medical sciences \cite{ref1,ref2,ref3,ref47,ref48,ref6,ref57}. It resolves the chemical specificity of elements from the resonance frequency of nuclear spins in order to obtain molecular structure or image of materials non-destructively. Conventional NMR, which often suffers from low detection efficiency, relies primarily on a large amount of sample molecules for signal accumulation \cite{ref78}. Many efforts have been provided to improve the sensitivity for example the hyper-polarization approaches \cite{hyperpolarization1,hyperpolarization,hyperpolarization2}, the magnetic resonance force microscopy\cite{mrfm} and optically-detected NMR \cite{odnmr,odnmr1}. Recently, the Nitrogen-Vacancy (NV) center in diamond has shown its outstanding capability as a nano-scale sensor for nuclear-spin detection that associates with a significantly-improved sensitivity even at the single-molecule level \cite{ref61,ref38}.

\clearpage
\begin{figure}[H]
\centering
\includegraphics[width=0.5\textwidth]{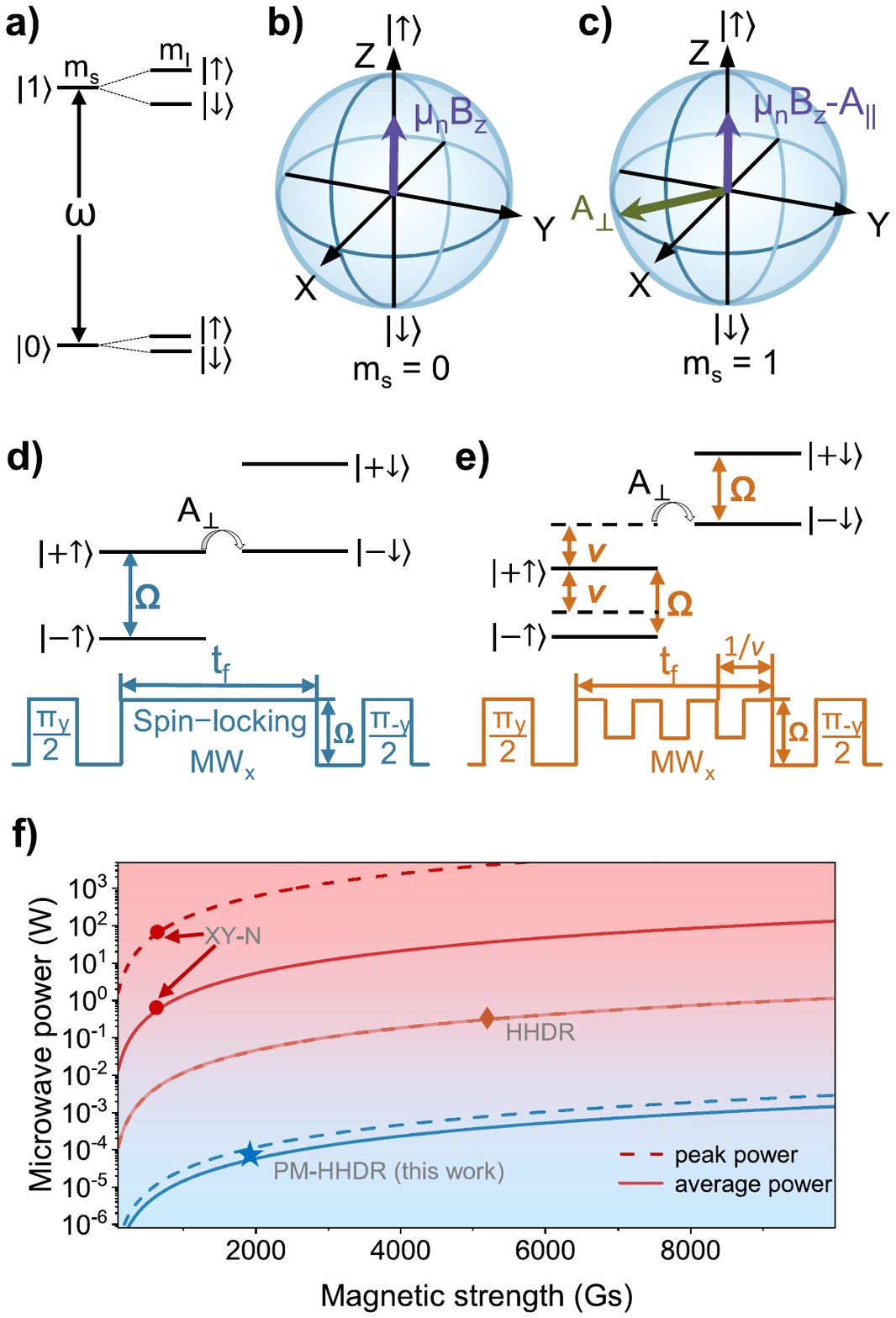}
\caption{ (Color online) \textbf{Principle description and main advantage of PM-HHDR}. \textbf{a)} Energy-level diagram of NV hybrid system consisted of an electron spin and a $^{13}$C nuclear spin. \textbf{b)-c)} The effective magnetic field on the nuclear spin when electron spin is in different subspaces. \textbf{d)-e)} (Upper) Energy level diagram in the electron-spin dressed-state space $|\pm\rangle$ when a resonant microwave field is applied for (d) HHDR and (e) PM-HHDR conditions. The coupling component $A_{\perp}$ induced the flip-flops between the $|+\uparrow\rangle$ and $|-\downarrow\rangle$ when the double resonance happens. (Lower) the pulse sequences for the corresponding control schemes. \textbf{f)} Comparison of microwave power used in different $^{13}$C nuclear-spin detection protocols. The peak (average) power is calculated from the Rabi frequency and microwave radiation efficiency in experiment \cite{omega} (see Supporting Information). The marker XY-N and HHDR in the plot represent experimental configurations in previous works respectively \cite{ref17,ref25}. 
} 
\label{fig1}
\end{figure}

In the field of NMR, the resonance frequency of nuclear spins is determined by the Larmor frequency and its local environment, \emph{i.e.}, the neighboring nuclear spins and surrounding electron distribution, which generates an additional field and changes the resonance frequency. The relative difference between the resonance frequency and the Larmor frequency also called the chemical shift, reveals fruitful knowledge of the host molecule \cite{ref24,ref54}. Therefore, a high spectral resolution is necessary for the NMR spectroscopy applications. Generally, the resolution is improved by increasing the external magnetic field which can enlarge the chemical shift. Besides, it also brings crucial benefits to the decoherence time and polarization of the sample spins under a high magnetic field.

NV-based NMR often utilizes the standard pulsed dynamical decoupling (pulsed-DD) sequences \cite{ref20,ref46,ref49,ref63}, such as XY-$N$ \cite{ref17,ref55} or Carr-Purcell-Meiboom-Gill-$N$ \cite{ref18,ref19,ref52}, as band-pass noise filters to detect the nuclear spins by matching their resonance frequency, as well as their extensions with a quantum(classical) memory to realize a high(arbitrary)-resolution signal detection\cite{arb_resolution,RN222}. But the pulsed-DD tends to work under low magnetic fields because the maximal band-pass frequency (limited by square root of the pulse power) has practical upper limitation and cannot increase as the magnetic field increases. An alternative scheme under high magnetic fields is the continuous-wave dynamical decoupling (CWDD) sequence \cite{ref74,zhouPRL124,kongPRL112}. It also requires the continuous microwave driving to match the resonance frequency that is linearly-scaled with the magnetic field. However, the microwave power has practical limitations and will induce various issues including power broadening of line-width and sample heating\cite{ref25}. A nanoscale NMR spectroscopy that can largely reduce the power requirements is highly benefited for high-field NMR and the related applications including bio-sensing applications \cite{ref79, ref80, ref81}.  

Recently, the phase-modulation (PM) technique is introduced into NV-based quantum sensing, which works as a mixer to down-convert high-frequency microwave fields to a proper range \cite{mixer1,mixer2}. Also, a theoretical method combining the PM and CWDD schemes has been proposed to reduce the power requirement for nuclear-spin detection under high magnetic fields \cite{ref27,power_limit}. In this work, we experimentally examine the PM-based CWDD scheme and realize the new scheme to detect \emph{in vivo} nuclear spin under a field up to 3015 Gs. The microwave power is about four orders of magnitude lower than the standard DD schemes. We show that the power-induced broadening of the spectrum is also suppressed, where four weakly-coupled $^{13}$C spins are distinguished from the spin bath with a minimal line-width of 2.1 kHz under 1840 Gs. Besides, we also contribute to the theoretical scheme by extending the resonant condition to both sidebands of the mixer, enabling its future applications in zero-field situations. This scheme can be easily extended to sensing of \emph{in vitro} nuclear spins or classical magnetic fields in general.

The  NV center in diamond has emerged as an excellent solid-state platform for quantum information processing \cite{ref34,ref5,ref36}. The center consists of a nitrogen atom with a neighboring vacancy and its negatively charged state NV$^-$ forms a spin triplet in the orbital ground state. The electron spin with nearby nuclear spins  form a hybrid system that has been used as quantum registers \cite{register1,register2,register3} for computation \cite{computation1,computation2,computation3,computation,computation5} and simulation \cite{simulation,simulation1} tasks, or interacting nodes of quantum network \cite{network1,network2,network3,network4}. NV center is also used as a nano-scale sensor to detect magnetic field \cite{ref7,ref8,ref58,lab2}, electric field \cite{ref9,ref10,ref65,ref67}, temperature \cite{ref11,ref12,ref71,ref72} or other spins \cite{ref15,ref39,ref56}. Under this situation, the Hamiltonian of the system is often formulated as
\begin{equation}\nonumber
\frac{H_0}{2\pi} =DS_z^2 -\gamma_e B_z S_z - \sum_{j}\gamma^{(j)}_{n}B_zI^{(j)}_{z}+\sum_{j}S_z\mathbf A^{(j)}\cdot  \mathbf I^{(j)}
\end{equation}
where $D\approx 2870$MHz is the zero-field splitting, $S_i$ ($I_i^{(j)}$) the electron ($j$-th nuclear) spin operator, $B_z$ the static magnetic field applied along the NV axis, $\gamma_e$ ($\gamma_n^{(j)}$) the gyro-magnetic ratio of electron ($j$-th nuclear) spin and $\mathbf A^{(j)}=(A^{(j)}_{z x}, A^{(j)}_{z y}, A^{(j)}_{z z})$ the interaction vector between electron and nuclear spin. In our experiment, only the subspace of $|m_s=0\rangle$ and $|m_s= 1\rangle$ of electron spin is utilized. A resonant microwave field with frequency $\omega$ is applied to drive the transition between $|0\rangle$ and $|1\rangle$ with a controlling Hamiltonian $H_{c}=\sqrt{2}\Omega\cos(\omega t+\phi)S^{(0,1)}_x$, where $\Omega$ is the Rabi frequency, $\phi$ the phase and $S^{(0,1)}_x=(|0\rangle\langle 1|+| 1\rangle\langle 0|)/\sqrt{2}$ the transition operator. In the following discussion, we consider a simple system that contains only the NV electron spin and a single $^{13}$C nuclear spin (the notation j is ignored in the following description).

Conventionally, the pulsed-DD scheme ($\emph{e.g.},$ XY-$N$) utilizes a basic operation sequence like $\tau-\pi-2\tau-\pi-\tau$, where $\tau$ is the interval time and $\pi$ pulse flips the electron spin between $|0\rangle$ and $|1\rangle$.  During this process, the extra phase generated by the nuclear spin is accumulated on the electron spin if its precessing speed ($2\gamma_n B_z+A_{zz}$) matches with $\tau$. This protocol requires the microwave to be as strong as possible to ensure $\pi$ pulse is much shorter than $\tau$. While in CWDD scheme (Fig.~\ref{fig1}d), the electron spin is instead prepared on one of the dressed states $|\pm\rangle\equiv(|0\rangle \pm |1\rangle)/\sqrt{2}$ and experiences a spin-locking process driven by another microwave. During this process, the spin would stay on the dressed state except for a global phase if no perturbation from the nuclear spin exists. Otherwise, when the electron-spin Rabi frequency $\Omega$ matches with the energy gap between nuclear-spin states, \emph{i.e.}, the Hartmann-Hahn (HH) condition is satisfied, the double resonance (DR) happens between states $|+\uparrow\rangle$ to $|-\downarrow\rangle$, where $|\uparrow\rangle$ and $|\downarrow\rangle$ are the nuclear-spin eigenstates. In the above protocols, the microwave field (power) is scaled linearly (quadratically) with the magnetic field $B_z$. As shown in Fig.~\ref{fig1}f, this makes them extremely hard to be applied under high magnetic fields.

\begin{figure*}[htb]
\centering
\includegraphics[width=1\textwidth]{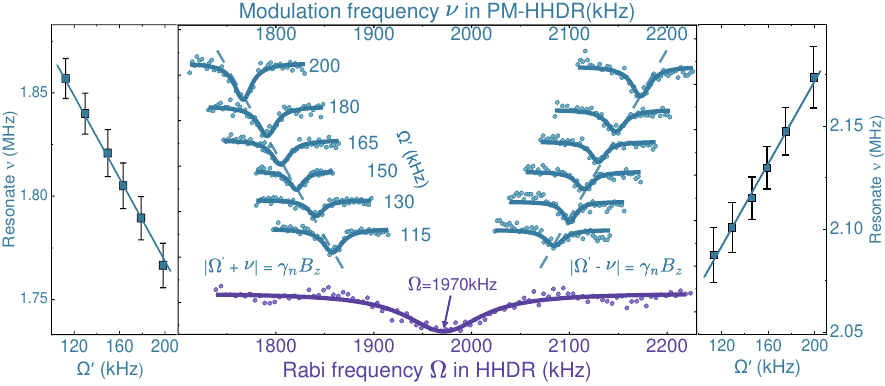}
\caption{ (Color online) \textbf{Experimental verification of resonant condition in the PM-HHDR and HHDR schemes}. The purple line is the conventional HHDR experiment results by varing the Rabi frequency $\Omega$. 
The blue spectrum refers to a PM-HHDR detection of $^{13} \mathrm{C}$ spin-bath using different effective Rabi frequency $\Omega^{'}$ ($\Omega^{'}=\Omega_{+}=\Omega_{-}$) and modulation frequency $\nu$ at 1840 Gs. The relation between the detected resonant position of $\nu$ and the effective Rabi frequency $\Omega^{'}$ is plotted in the left (right) column with the fitted slope -1.046$\pm0.15$ (1.006$\pm0.19$), matching well with Eq.\ref{HHDR_condition}. The dip positions in the center column are placed along the dashed lines in parallel with the fitted slops. The experiment is performed on NV1.
}
\label{fig2}
\end{figure*}

Recently, new schemes based on HHDR are proposed where the phase or amplitude of the microwave field is modulated in a regular \cite{ref27} or arbitrary\cite{power_limit} way to compensate for the mismatch between Rabi frequency and nuclear-spin energy gap. To elaborate this new protocol, we first explain the original HHDR and show its schematic in Fig.~\ref{fig1}b. In electron-spin $m_s=0$ subspace, the hyperfine interaction takes no effect and the nuclear spin precesses under magnetic field $B_{z}$ with frequency $\gamma_nB_z$ only. But in $m_s=1$ subspace (Fig.~\ref{fig1}c) it generates an extra precessing $A_{\parallel}$ along the $z$ axis and an effective field $A_{\perp}/\gamma_n$ in the $x$-$y$ plane on nuclear spin, where $A_{\parallel}=A_{zz}$ and $A_{\perp}=\sqrt{A_{zx}^2 + A_{zy}^2}$. As a result, when electron spin rotates along the $x$-axis during the spin-locking, the nuclear spin experience a $z$-axis precessing with frequency $\gamma_{n} B_{z}-\frac{1}{2} A_{\parallel}$ in average and an effective microwave field in $x$-$y$ plane with frequency $\Omega$. Once the HH condition is satisfied, \emph{i.e.},  $\Omega=\gamma_{n} B_{z}-\frac{1}{2} A_{\parallel}$, the nuclear-spin resonance happens and the electron-spin state is changed by the hyperfine interaction as well.

In the phase-modulated (PM-) HHDR scheme (Fig.~\ref{fig1}e) , the driven field is replaced with 
\begin{equation}
H_{c}= \sqrt{2}\Omega_{+} S^{(0,1)}_x \cos (\omega t)+\sqrt{2} \Omega_{-} S^{(0,1)}_x \cos (\omega t-\phi),
\end{equation}
where the phase $\phi$ switches periodically between the values 0 and $\pi$ with a modulation frequency $\nu$. The total effect of this modulation is to change the electron-spin rotating speed between $(\Omega_{+}+\Omega_{-})$ and $(\Omega_{+}-\Omega_{-})$ in the same frequency $\nu$. In experiment, we make  $\Omega^{'} =\Omega_{+}=\Omega_{-}$ to ensure a best signal contrast (see Supporting Information).
 The effective magnetic field on the nuclear spin is also modulated to generate two sidebands with frequency $\Omega^{'}+\nu$ and $\Omega^{'}-\nu$. Then the resonant condition in the PM-HHDR scheme is changed to
\begin{equation}
\label{HHDR_condition}
\left|\Omega^{'} \pm \nu\right|=\left|\gamma_{n} B_{z}-A_{\parallel}/2\right|.
\end{equation}
and the signal in the resonance $\nu$ is described by $\cos^2\left( A_{\perp} J_1(4 \Omega^{'} / \pi \nu) t_f/ 4 \right)$, where $\Omega^{'}$ is defined as effective Rabi frequency, $t_f$ is integration time, $J_1$ is the first kind of Bessel function. This new condition brings a possibility to make the microwave power significantly reduced when we drive the electron spin to achieve the double resonance under high magnetic fields (Fig.~\ref{fig1}f). Note that $J_1$ associates with a relatively small value in experiment, which weakens the effective coupling between the sensor and target spins, thus reduces the sensitivity of this scheme. It is worth pointing out that the theoretical proposal \cite{ref27} only consider the single-side resonant situation $\Omega^{'} + \nu=\gamma_{n} B_{z}-A_{\parallel}/2$, which is different with Eq.\ref{HHDR_condition} here (see Supporting Information).

Here, we utilize single NV centers (NV1 and NV2) in diamond to detect intrinsic $^{13}$C nuclear spins at 1840 Gs as an example. The PM-HHDR spectrum is obtained by sweeping the modulation frequency $\nu$ with $\Omega^{'}$ fixed. The result is shown in Fig.\ref{fig2}, where we can observe two resonance dips symmetrical about $\gamma_nB_z$, \emph{i.e.}, the Larmor frequency of $^{13}$C nuclear spin. Additionally, with decreasing $\Omega^{'}$ the positions of resonance moves towards the Larmor frequency. The relation between the resonance position and $\Omega^{'}$ is highlighted in the Fig.~\ref{fig2}, which matches well with the condition defined in Eq.\ref{HHDR_condition}. Most importantly, the effective Rabi frequency $\Omega^{'}$ is reduced to around 1/20 (1/250) of the microwave field used in conventional HHDR (XY-N) where the modulation frequency $\nu$ is chosen much larger than $\Omega^{'}$ (see Supporting Information). Noted that $A_{\parallel}$ is averaged for different $^{13}$C nuclear spins, thus it is not apparently shown in the relation. 

Since the resolution of the HHDR spectrum is mainly determined by the noise in microwave amplitude, low-power control can also improve the spectrum in principle \cite{ref59}. However, this also weaken the decoupling efficiency of the CWDD sequence that protects the electron spin against ($z$-axis) magnetic fluctuations. In order to improve the performance, a single-spin lock-in detection \cite{2201.06002,ref33} is realized to monitor the electron-spin resonance in real-time together with a PID-based temperature control of the setup within $\pm5$ mK.

\begin{figure*}[htb]
\centering
\includegraphics[width=0.8\textwidth]{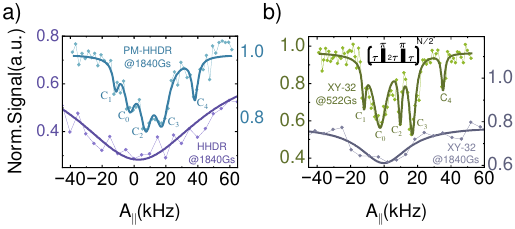}
\caption{ (Color online) \textbf{Comparison of nuclear spin detection spectrums} \textbf{a)} The upper line shows the high-resolution PM-HHDR spectrum for ${}^{13}C$ nuclear spins measured using interrogation time $t_f$ = 300 $\mu s$ and Rabi frequency $\Omega^{'}=104$ kHz ($\Omega^{'}=\Omega_{+}=\Omega_{-}$) under a magnetic field at 1840 Gs. The coupling between electron and nuclear spins is extracted from the spectrum in Tab.\ref{tab1}. The lower line shows the conventional HHDR spectrum under the same magnetic field using $t_f$ = 8 $\mu s$ and $\Omega\sim1970$ kHz (the Larmor frequency of the nuclear spin). \textbf{b)}  XY-32 spectrum under a magnetic field at 522 Gs (upper) and 1840 Gs (lower) using an interrogation time of 640 $\mu s$. The experiment is performed on NV1. }
\label{fig:details}
\end{figure*}

Meanwhile, we optimize the experimental parameters, and finally the single resonance dip is split into five narrow lines as shown in Fig.~\ref{fig:details}a (the upper line). By fitting the spectrum, four weakly-coupled individual $^{13}$C nuclear spins and a spin-bath signal are assigned to each dip. The calculated coupling parameter is listed in Tab.\ref{tab1}. Because of the reduction in microwave noise, a minimal (average) line-width of 2.1 (3.5) kHz is achieved. For comparison, we realize conventional HHDR spectroscopy under the same field in Fig.~\ref{fig:details}a (the lower line). Since the microwave amplitude is around 20 times higher than that is used in PM-HHDR, the line-width is over 100 kHz and none of the weakly-coupled nuclear spins can be observed (see Supporting Information).

\begin{table}
    \caption{Hyperfine interaction parameters $A_{\parallel}$ of four $^{13}$C nuclear spins from PM-HHDR and XY-32 experiments on NV1, the number in braket is the corresponding line-width. $C_0$ refers to the spin-bath signal.}
    \label{tab1}
        \begin{tabular}{cccc}
            Spin No.& $A_{\parallel}$ (PM-HHDR) &   $A_{\parallel}$ (XY-32) & Unit  \\
            \hline
            $\rm{C_0}$ & -2.4 (4.1) & -2.3 (13.4) & kHz  \\
            $\rm{C_1}$ &-11.3 (2.1) &  -11.8 (2.3) &kHz \\
            $\rm{C_2}$ & 7.0 (4.5) &9.8 (2.6)  &kHz  \\
            $\rm{C_3}$ & 17.2 (5.3) &  17.0 (4.7) &kHz \\
            $\rm{C_4}$ & 38.0 (2.2) & 35.3 (2.4) &  kHz\\
        \end{tabular}
\end{table}

\begin{figure}[tb]
    \centering
    \includegraphics[width=0.5\textwidth]{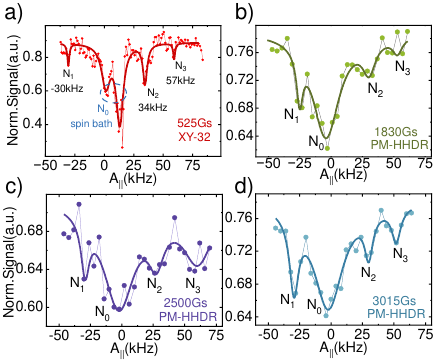}
    \caption{ (Color online) \textbf{The PM-HHDR ${}^{13}C$ nuclear spins spectral under different magnetic fields.} The figure \textbf{a)} is the XY-32 spectrum under the magnetic field at 525 Gs. We can clearly identify three ${}^{13}C$ nuclear spins ($N_{1},N_{2},N_{3}$) and spin-bath signal ($N_{0}$). The figure \textbf{b)}\textbf{c)}\textbf{d)} are the spectral of the nuclear spin under the magnetic fields of 1830 Gs, 2500 Gs, and 3015 Gs, respectively (the interrogation time $t_f$ = 300 $\mu s$, the Rabi frequency $\Omega^{'}=\Omega_{+}=\Omega_{-}$ is 150 kHz, 170 kHz, 200 kHz). The value of the hyperfine interaction parameters $A_{\parallel}$ be detected is -29 kHz,31 kHz,52 kHz and the spin-bath (10 kHz) signal. The experiment is performed on NV2. }
    \label{fig4}
\end{figure}

To verify this result, we perform a standard XY-32 experiment on the same NV center. As shown in Fig.~\ref{fig:details}b (the upper line), a spectrum with five distinguishable dips is observed under a field near 500 Gs and the minimal (average) line-width in the spectrum is 2.3 (3.0) kHz respectively. The extracted coupling information (also shown in Tab.\ref{tab1}) matches well with the result from PM-HHDR. The experiment under a higher magnetic field (the same field as PM-HHDR) is also realized, where the spectral lines become overlapped and no splitting can be observed anymore (the lower line in Fig.~\ref{fig:details}b). For even higher magnetic fields, the scheme works well and the results are shown in Fig.~\ref{fig4}. Due to the maximal sampling frequency of the microwave pulse generator, the frequency-modulation resolution is increased to 2 kHz and we have to boost the Rabi frequency $\Omega^{'}$ to broaden the line-width accordingly (see Supporting Information).

Here we demonstrate a nano-scale nuclear-spin detection spectroscopy under high magnetic fields and achieve an spectral line-width down to 3.5 kHz at about 1840 Gs. The microwave power is about two orders below previous double resonance schemes, which associates with a detection bandwidth over 100 kHz (see Supporting Information). The performance is mainly limited by the minimal configuration precision of the modulation frequency $\nu$ in our case. High-resolved pulse control based on time-delay lines can further improve the line-width to around 10 Hz. 
In general cases, the ultimate limit is defined by the rotating-frame relaxation time $T_{1}^{\rho}$ of the electron spin, where $T_{1}^{\rho}$ is measured to be over mili-second in the experiment thus the spectral line-width is possibly pushed down to the sub-kHz level in the future. Although our work is focused on the electron-nuclear hyperfine interactions, this method can also applied to more general spin-based quantum sensing fields, for example, detection of electron-electron spin interaction \cite{EE}, chemical shift analysis \cite{RN221} and spin manipulation in two-dimensional materials\cite{ref101}. Finally, we have extended the resonant condition to both sidebands towards the real resonance \cite{ref27}, which could potentially be utilized in the ultra-low or zero-field cases, \emph{i.e.}, to up-convert the required Rabi frequency to a proper range in the future \cite{ref23}.

\begin{acknowledgement}

The authors thank Bing Chen and Ying Dong for helpful discussion. This work was supported by the National Natural Science Foundation of China (Grant Nos. 92265114, 92265204), the Fundamental Research Funds for the Central Universities (Grant No. 226-2023-00139),the Innovation Program for Quantum Science and Technology (Grant No. 2021ZD0302200).

\end{acknowledgement}

\begin{suppinfo}
\begin{itemize}
  \item Supporting Information: Details of the setup and experimental scheme, and analysis of the performance. (PDF).
\end{itemize}

\end{suppinfo}
\providecommand{\latin}[1]{#1}
\makeatletter
\providecommand{\doi}
  {\begingroup\let\do\@makeother\dospecials
  \catcode`\{=1 \catcode`\}=2 \doi@aux}
\providecommand{\doi@aux}[1]{\endgroup\texttt{#1}}
\makeatother
\providecommand*\mcitethebibliography{\thebibliography}
\csname @ifundefined\endcsname{endmcitethebibliography}
  {\let\endmcitethebibliography\endthebibliography}{}


\end{document}